\documentstyle[twoside,fleqn,espcrc2]{article}
\newcommand{\stru}[1]{\rule{0ex}{#1}}   % for vspace in formulae and table
\newcommand{\str}{\rule{0ex}{2.7ex}}    % strut, to make a line higher

\newcommand{\half}{{1\over2}}		% One half

\newcommand{\subs}[1]{{\mbox{{\protect\scriptsize #1}}}}
\newcommand{\U}{\tilde{u}}
\title{
  \quad\vskip-3.3cm \hfill {\normalsize
  \begin{tabular}[t]{l}
                     \rule{0ex}{1ex}hep-lat/9211047 \\[.5ex]
                     \rule{0ex}{1ex}FSU-SCRI-92-165 \\[.4ex]
                     \rule{0ex}{1ex}November 1992
  \end{tabular}}
  \vskip1.0cm  % 1.1 already makes another page.
The Loop-Cluster Algorithm for the Case of the 6 Vertex Model%%
   \thanks  {To appear in {\em Lattice '92}, Amsterdam 1992,
             ed.\ J. Smit et al.,
             Nucl.\ Phys.\ B (Proc.\ Suppl.).}
}%end of \title
\author{%
Hans Gerd Evertz\address{%
Supercomputer Computations Research Institute,
Florida State Univ., Tallahassee, FL 32306, USA}
and  Mihai Marcu\address{%
School of Physics and Astronomy,
Raymond and Beverly Sackler Faculty of Exact Sciences,
Tel Aviv University, 69978 Tel Aviv, Israel}}
\begin{document}
\begin{abstract}
We present the loop algorithm,  a new type of cluster
algorithm that we recently introduced for the F~model. Using the framework of
Kandel and Domany, we show how to generalize the algorithm to the arrow
flip symmetric 6 vertex model.  We propose the principle of least
possible freezing as the guide to choosing the values of free
parameters in the algorithm. Finally, we briefly
discuss the application of our algorithm to simulations of quantum spin
systems. In particular, all necessary information is provided for
the simulation of spin $\half$ Heisenberg and $xxz$ models.
\end{abstract}
\maketitle

\section{INTRODUCTION}

Cluster algorithms, originally introduced for the Ising model \cite{SW}
and then generalized to various other situations
\cite{ClusterReviews,KandelDomany}, are one of the promising ways of
overcoming critical slowing down. Recently \cite{BCSOScluster,LoopAlg}
we introduced algorithms for
vertex models \cite{Baxter,ReviewLieb},  which are  the first cluster
algorithms for models with constraints. While \cite{BCSOScluster} is an
adaptation of an algorithm originally devised for solid-on-solid models,
the loop algorithm introduced in \cite{LoopAlg} does not resemble any
existing scheme.

In \cite{LoopAlg} we presented the loop algorithm for the F~model. This
enabled us to present our idea as clearly as possible.  Here we shall
show how to generalize it for the 6 vertex model, which has an
additional coupling and a richer phase structure (see below). The
framework of \cite{KandelDomany} proves to be an extremely useful
tool here.

Our scheme is devised such as to take into account the constraints
automatically, and to allow a simple way to construct the  clusters.
After defining the relevant probabilities, we find that we still have
some free parameters. In order to optimize the algorithm, we introduce
the principle of minimal freezing. As seen from the example of the
F~model, this choice of parameters is of utmost importance.

The loop algorithm can be further generalized to more complicated vertex
models. However, as it stands, there are already important applications:
quantum spin systems can be simulated by mapping them to vertex models
\cite{QMC}. In particular, the loop algorithm for the 6 vertex model
presented here can be used to simulate spin $\frac{1}{2}$ Heisenberg
ferromagnets and antiferromagnets, and $xxz$ models, even in more than
one dimension.

\section{THE 6 VERTEX MODEL}

The six vertex model \cite{Baxter,ReviewLieb} is defined on a square
lattice. On the bonds there lives an Ising-like variable that is usually
represented as an arrow. For example, arrow up or right means plus one,
arrow down or left means minus one. At each vertex, there are two
incoming and two outgoing arrows. In fig.~1 %%\ref{fig1}
we show the six
possible configurations at a vertex, numbered as in
\cite{Baxter,ReviewLieb}.

%
%%%%%%% PICTURE OF VERTICES %%%%%%%%%%%%%%%%%%%%%%%%%%%%%%%%%%%%
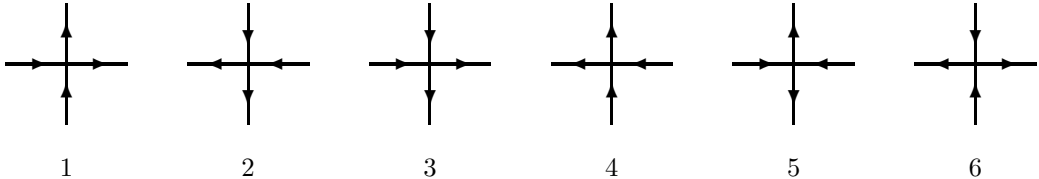
\begin{figure*}[t] \label{fig1}
\begin{center}
\setlength{\unitlength}{.0005\textwidth}
%\begin{picture}(1700,300)(0,20)  % would be "correct"
\begin{picture}(1700,200)(0,90) % 200: higher; 90: caption closer
\thicklines
%
%%% DEFINE ARROWS (length 100) and VERTEX
\newsavebox{\RIGHT}
\newsavebox{\LEFT}
\newsavebox{\UP}
\newsavebox{\DOWN}
\sbox{\RIGHT}{\put(  0,0  ){\vector( 1, 0){70}}\put(70,0 ){\line(1,0){30}}}
\sbox{\UP}   {\put(  0,0  ){\vector( 0, 1){70}}\put(0 ,70){\line(0,1){30}}}
\sbox{\LEFT} {\put(100,0  ){\vector(-1, 0){70}}\put(0 ,0 ){\line(1,0){30}}}
\sbox{\DOWN} {\put(  0,100){\vector( 0,-1){70}}\put(0 ,0 ){\line(0,1){30}}}
\newcommand{\VERTEX}[5]{\begin{picture}(0,0)
                              \put(-100,0   ){\usebox{#1}}       %left
                              \put(   0,0   ){\usebox{#2}}       %right
                              \put(   0,-100){\usebox{#3}}       %bottom
                              \put(   0,0   ){\usebox{#4}}       %top
                              \put(   0,-170){\makebox(0,0){#5}} %text
                        \end{picture}}
%%% NOW DRAW 6 VERTICES
%
\put( 100,200){\VERTEX{\RIGHT}{\RIGHT}{\UP}  {\UP}  {1}}
\put( 400,200){\VERTEX{\LEFT} {\LEFT} {\DOWN}{\DOWN}{2}}
\put( 700,200){\VERTEX{\RIGHT}{\RIGHT}{\DOWN}{\DOWN}{3}}
\put(1000,200){\VERTEX{\LEFT} {\LEFT} {\UP}  {\UP}  {4}}
\put(1300,200){\VERTEX{\RIGHT}{\LEFT} {\DOWN}{\UP}  {5}}
\put(1600,200){\VERTEX{\LEFT} {\RIGHT}{\UP}  {\DOWN}{6}}
\end{picture}
\caption[fig1]{\parbox[t]{.80\textwidth}{
               The six vertex configurations, $u=1,...,6$
               (using the standard conventions of \cite{Baxter}).}}
\end{center}
\end{figure*}
%%%%%%%%%%%%%% END OF PICTURE %%%%%%%%%%%%%%%%%%%%%%%%%%%%%%%%%%%
%

The statistical weight of a configuration is given by the product over
all vertices of the vertex weights $\rho(u)$.
For each vertex there are 6 possible weights $\rho(u)$, $u=1,...,6$.
We assume the vertex
weights to be symmetric under reversal of all arrows. So in standard
notation \cite{Baxter,ReviewLieb} we have:
\begin{equation}\label{e1}
\begin{array}{l}
  \rho(1)  =  \rho(2)  =  a \; , \\
  \rho(3)  =  \rho(4)  =  b \; , \\
  \rho(5)  =  \rho(6)  =  c \; .
\end{array}
\end{equation}

The six vertex model basically has two types of phase transitions: of
Kosterlitz-Thouless type and of KDP type \cite{Baxter,ReviewLieb}.
A sub-model exhibiting the former is the F~model, defined by
$c=1$, $a=b=\exp{(-K)}$, $K\geq0$.
For the latter transition an example is the KDP
model itself, defined by $a=1$, $b=c=\exp{(-K)}$, $K\geq0$.

\section{THE LOOP ALGORITHM}

If we regard the arrows on bonds as a vector field, the constraint at
the vertices is a zero-divergence condition. Therefore every
configuration change can be obtained as a sequence of {\em loop-flips}.
%operations defined as follows.
By ``loop'' we denote an oriented, closed,
non-branching (but possibly self-intersecting)
path of bonds, such that all arrows along the path point in the
direction of the path. A loop-flip reverses the direction  of
%(``flips'')
all arrows along the loop.

Our cluster algorithm performs precisely such operations, with
appropriate probabilities.
It constructs closed paths consisting of one or
several loops without common bonds. All loops in this path are flipped
together.

We shall construct the path iteratively, following the direction of the
arrows. Let the bond $b$ be the latest addition to the path. The arrow on
$b$ points to a new vertex $v$. There are two outgoing arrows at $v$,
and what we need is a unique prescription for continuing the path
through $v$. This is provided by a {\em break-up} of the vertex $v$.
In addition to the break-up, we have to allow for {\em freezing} of $v$.
By choosing suitable probabilities for break-up and freezing we shall
satisfy detailed balance.

The {\em break-up} operation is defined by splitting $v$ into two
pieces, as shown in fig.~2. %%\ref{fig2}.
The two pieces are either two
corners or two straight lines. On each piece, one of the arrows points
towards $v$, while the other one points away from $v$.
Thus we will not allow e.g.\ the ul--lr
break-up for a vertex in the configuration 3.
If we break up $v$, the possible new configurations are obtained by
flipping (i.e.\ reversing both arrows of) the two pieces independently.
On the other hand, if we freeze $v$, the only possible configuration
change is to flip all four arrows.

%
%%%%%%% PICTURE OF BREAK-UP OF VERTICES AND OF STRAIGHT LINES %%%%%%%%%%%%
\begin{figure}[tb] \label{fig2}
\begin{center}
\setlength{\unitlength}{.0004\textwidth} % to fit in one column
%\begin{picture}(1020,310)(0,20) % would be "correct"
\begin{picture}(1020,210)(0,80) % 260: move higher; 80: caption closer
\thicklines
\put(100,190){\line(-1, 0){100}}
\put(100,190){\line( 0,-1){100}}
\put(120,210){\line( 1, 0){100}}
\put(120,210){\line( 0, 1){100}}
\put(115,20){\makebox(0,0){ll--ur}}
\put(500,210){\line(-1, 0){100}}
\put(500,210){\line( 0, 1){100}}
\put(520,190){\line( 1, 0){100}}
\put(520,190){\line( 0,-1){100}}
\put(510,20){\makebox(0,0){ul--lr}}
\put(910,200){\line(-1, 0){110}}  % left
\put(910,200){\line( 1, 0){110}}  % right
\put(910,180){\line( 0,-1){ 90}}  % down
\put(910,220){\line( 0, 1){ 90}}  % up
\put(910,200){\oval(40,40)[l]}    % half-circle, left
\put(910,20){\makebox(0,0){straight}}
\end{picture}
\caption[fig2]{ %%\parbox[t]{.80\hsize}{
                The three break-ups of a vertex:
                ll--ur (lower-left--upper-right),
                ul--lr (upper-left--lower-right),
                and straight.
              } %%}
\end{center}
\vskip-1.5ex   % less space between figure and text
\end{figure}
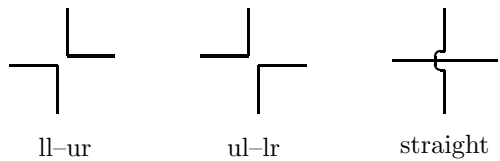
%
%%%%%%%%%%%%% END OF PICTURE %%%%%%%%%%%%%%%%%%%%%%%%%%%%%%%%%%%%%%%%%
%

The break-up and freeze probabilities are conveniently described within
the general framework for cluster algorithms proposed by Kandel and
Domany \cite{KandelDomany}. It is sufficient to give them for one
vertex, which is in the current configuration $u$.
We define 6 new interactions (weight functions) $\rho_i$, $i=1,...,6$,
corresponding to specific break-up and freeze operations.
(The labelling of the new interactions is completely
arbitrary, and the fact that we have six of them is just a coincidence).
For each vertex in configuration $u$, we replace
with probability $p_i(u)$ the original interaction $\rho$ by
the new interaction $\rho_i$.
Detailed balance and the proper normalization of probabilities require
that for every $u$
\begin{equation}\label{e2}
p_i(u) = q_i \frac{\rho_i(u)}{\rho(u)} \; , \quad \sum_i p_i(u) = 1 \; ,
\end{equation}
where $q_i \!\geq\! 0$ are parameters.

%%%%%%%%%%%% TABLE 1 %%%%%%%%%%%%%%%%%%%%%%%%%%%%%%%%%%%%%%%%
\begin{table*}[t]    \label{tab1}
%% \centering
\begin{tabular}{|c|c|c|c|c|c|c|}
\hline\str
 $i$ & 1 & 2 & 3 & 4 & 5 & 6 \\
 action & freeze 1,2 & freeze 3,4 & freeze 5,6 & ll--ur & ul--lr & straight
\\[.5ex]\hline \stru{4ex}
 $\rho_i(\U)$
& \parbox[c]{20mm}{1, $\rho(\U)\!=\!a$ \\ 0, else}
& \parbox[c]{20mm}{1, $\rho(\U)\!=\!b$ \\ 0, else}
& \parbox[c]{20mm}{1, $\rho(\U)\!=\!c$ \\ 0, else}
& \parbox[c]{20mm}{0, $\rho(\U)\!=\!a$ \\ 1, else}
& \parbox[c]{20mm}{0, $\rho(\U)\!=\!b$ \\ 1, else}
& \parbox[c]{20mm}{0, $\rho(\U)\!=\!c$ \\ 1, else}
\\[2.0ex]\hline \stru{4ex}
 $p_i(u)$
& \parbox[c]{21mm}{$q_1/a$, $\rho(u)\!=\!a$ \\ 0, else}
& \parbox[c]{20mm}{$q_2/b$, $\rho(u)\!=\!b$ \\ 0, else}
& \parbox[c]{20mm}{$q_3/c$, $\rho(u)\!=\!c$ \\ 0, else}
& \parbox[c]{20mm}{0, $\rho(u)\!=\!a$ \\ $q_4/\rho(u)$, else}
& \parbox[c]{20mm}{0, $\rho(u)\!=\!b$ \\ $q_5/\rho(u)$, else}
& \parbox[c]{20mm}{0, $\rho(u)\!=\!c$ \\ $q_6/\rho(u)$, else}
\\[2.0ex] \hline
\end{tabular}
%
%%! in espcrc2, a table caption produces a nasty linebreak after table-#.
%%! \caption[dummy]{ %%\parbox[t]{0.70\textwidth}{
\vskip1ex  Table~1.  %%\ref{tab1}.
\parbox[t]{0.9\hsize}{
                  New interactions $\rho_i(\U)$
                  and the probabilities $p_i(u)$
                  to choose them at a vertex in current configuration~$u$.
                  See eq.\ (\ref{e2}).
                } %%}
\end{table*}
As discussed in \cite{KandelDomany} (see also table~1),  %%\ref{tab1}),
{\em freezing} is described by introducing one new interaction for each
different value of $\rho(u)$.
For example, to freeze the value $a$, we choose the
interaction $\rho_1$ to be
$\rho_1(u)=1$ if $\rho(u)=a$, and $\rho_1(u)=0$ otherwise.
In other words, if $u$ is 1 or 2, the Boltzmann weight $\rho_1(u)$ is one,
so transitions between 1 and 2 cost nothing; the vertex configurations 3,
4, 5, and 6 are however not allowed with $\rho_1$.

Each {\em break-up} is also described by one new interaction. As an example
take the ul--lr break-up. It is given by the new interaction number five,
with $\rho_5(u) = 1$ if $\rho(u)=a$ or $c$,
and $\rho_5(u) = 0$ if $\rho(u)=b$.
In other words, with the new interaction $\rho_5$, transitions
between 1, 2, 5 and 6 cost nothing, while the vertex configurations 3
and 4 are not allowed.
This corresponds precisely to allowing independent corner flips
in a ul--lr break-up (see figs.~1,2).  %%\ref{fig1},\ref{fig2}).

The full list of new weights $\rho_i(u)$ and probabilities
$p_i(u)$ to choose them are given in table~1.  %%\ref{tab1}.
{}From (\ref{e2}) we also obtain:
\begin{equation}\label{e3}
\begin{array}{l}
q_1+q_5+q_6=a \; , \\
q_2+q_4+q_6=b \; , \\
q_3+q_4+q_5=c \; .
\end{array}
\end{equation}

Assume now that we have broken or frozen all vertices. Starting from a
bond $b_0$, we proceed to construct a closed path by moving in the arrow
direction. As we move from vertex to vertex, we always have a unique way
to continue the path. At broken vertices the path enters the vertex
through one bond and leaves it through another.
If the last bond $b$ added to the cluster points to a frozen vertex $v$,
the path bifurcates in the directions of the two outgoing arrows of $v$.
One of these directions can be considered as
belonging to the loop we came from, the other one as belonging
to a new loop. Since we also have to flip the second incoming
arrow of $v$, we are assured that this new loop also closes.
The two loops have to be flipped together.
In general, the zero-divergence condition
guarantees that all loops will eventually close.

We have now finished describing the procedure for constructing
clusters. In order to specify the algorithm completely, we must choose
values for the constants $q_i$, and decide how the clusters are flipped.
The former problem is of utmost importance, and it is the object of the
next chapter. For the cluster flips, we may use
both the Swendsen-Wang procedure and the single cluster flip
\cite{ClusterReviews}.
In \cite{LoopAlg} we used the latter, and obtained a drastic reduction
of critical slowing down.

\section{OPTIMIZATION}

We have seen that freezing forces loops to be flipped together.
%This effect is somewhat reminiscent of the problems caused by
%frustration in other cluster algorithms \cite{ClusterReviews},
%where
%We would certainly like large, fractal clusters, but on the other
%hand
Previous experience with cluster algorithms \cite{ClusterReviews}
suggests that it is advantageous to be able to flip them independently.
We therefore introduce the principle of {\em minimal freezing}
as a guide for choosing the constants $q_i$:
we shall minimize the freezing probabilities,
given the constraints (\ref{e3}) and $q_i \geq 0$.
%Furthermore, the larger a weight $\rho(u)$ is, the more we shall try to
%avoid its freezing.
In \cite{LoopAlg} we report that for the case of the F model, optimization
by minimal freezing does indeed minimize critical slowing down.
Here we discuss optimization for the 4 phases of the 6 vertex model,
usually denoted by capital roman numerals \cite{Baxter,ReviewLieb}.

Let us first look at phase~IV, where $c > a+b$.
To minimize the freezing of weight $c$ we have to minimize $q_3$.
{}From (\ref{e3}),  $q_3 = c-a-b + q_1+q_2+2q_6$.
With $q_i \geq 0$ this implies $q_\subs{3,min} = c-a-b$.
The minimal value of $q_3$ can  only be chosen if {\em at the same time}
we set $q_1=q_2=0$, i.e.\ minimize
 (in this case do not allow for)
the freezing of the smaller weights $a$ and $b$.
The optimized parameters for phase~IV are then:
\begin{equation} \label{e4}
\begin{array}{lll}
q_1=0, & q_2=0, & q_3=c-a-b, \\
q_4=b, & q_5=a, & q_6=0 \;.
\end{array}
\end{equation}

In phase I the situation is technically similar.
Here $a > b+c$, and we minimize freezing with $q_1=a-b-c$ and $q_2=q_3=0$.
The same holds for phase II, $b>a+c$, where we obtain minimal freezing
for $q_2=b-a-c$ and $q_1=q_3=0$.

Phase III (the massless phase) is characterized by
$a,b,c$ $< \half (a+b+c)$.
Here we can set all freezing probabilities to zero.
Thus,
\begin{equation} \label{e5}
\begin{array}{lll}
q_1=0, & 2q_4=b+c-a \; , \\
q_2=0, & 2q_5=c+a-b \; , \\
q_3=0, & 2q_6=a+b-c \;.
\end{array}
\end{equation}

The F~model is obtained from (\ref{e4}) and (\ref{e5})
as the special case $a=b \leq 1$, $c=1$. One can easily
see that for this case we recover the discussion of \cite{LoopAlg}.
(Notice that since $a=b$, in the F~model the straight break-up
will be called freezing).

\section{APPLICATIONS, CONCLUSIONS}

We have presented a new type of cluster algorithm, the loop algorithm, for
the case of the six vertex model. For the F~model, the algorithm has been
shown in \cite{LoopAlg} to beat critical slowing down.

Particularly promising is the possibility of
{\em accelerating Quantum Monte Carlo simulations} \cite{QMC,QMCpaper}.
Quantum spin systems in one and two dimensions can be mapped into
vertex models in $1+1$ and $2+1$ dimensions via the Trotter formula
and suitable splittings of the Hamiltonian \cite{QMC}.
The simplest example is the spin $\half$ $xxz$ quantum chain, which is
mapped directly into the 6-vertex model.
For higher spins, more complicated
vertex models result (e.g.\ 19-vertex model for spin one).

For $(2+1)$ dimensions, different splittings of the Hamiltonian can
lead to geometrically quite different situations
\cite{QMC,QMCpaper}.
We can e.g.\ choose between 6-vertex models on a complicated $2+1$
dimensional lattice, and models on a {\em bcc} lattice with 8
bonds (and a large number of configurations) per vertex.
Notice that
for the simulation of the 2-dimensional Heisenberg antiferromagnet
using the former splitting, all relevant formulas have been worked out
in the present paper.

\section*{ACKNOWLEDGEMENTS}

This work was supported in part by the Ger\-man-Israeli
Foundation for Research and Development (GIF) and
by the Basic Research Foundation of
the Israel Academy of Sciences and Humanities.
We would like to express our gratitude
to the HLRZ at KFA J\"ulich % where most of our computer runs were performed.
and to the DOE for providing the necessary computer time
for the F~model study.

\end{document}